\documentclass[longbibliography,physrev,aps,twocolumn]{revtex4-2}

\usepackage[T1]{fontenc}
\usepackage{lmodern}
\usepackage{microtype}
\usepackage{mathtools}
\usepackage{amssymb}
\usepackage{bm}
\usepackage{hyperref}
\hypersetup{
  colorlinks=true,
  linkcolor=blue,
  citecolor=blue,
  urlcolor=blue,
  pdftitle={Rank-One Loss Maximizes the Spectral Response Strength of Passive Full-Order Exceptional Points}
}

\usepackage{dsfont} % \mathds

\newcommand{\id}{\mathds{1}}
\newcommand{\ii}{\mathrm{i}}
\newcommand{\EP}{E_{\mathrm{EP}}}
\newcommand{\Ham}{\hat{H}}
\newcommand{\Herm}{\hat{K}}
\newcommand{\Loss}{\hat{\Gamma}}
\newcommand{\Nil}{\hat{N}}
\newcommand{\Green}{\hat{G}}
\newcommand{\Pert}{\hat{V}}
\newcommand{\SchurT}{\hat{T}}
\newcommand{\Corr}{\hat{C}}
\newcommand{\Unitary}{\hat{U}}
\newcommand{\tr}{\operatorname{Tr}}
\newcommand{\rank}{\operatorname{rank}}
\newcommand{\im}{\operatorname{Im}}
\newcommand{\norm}[1]{||{#1}||}
\newcommand{\mel}[3]{\left\langle #1\middle|#2\middle|#3\right\rangle}

\newcommand{\HL}[1]{#1}

\begin{document}

\title{Rank-One Loss Maximizes the Spectral Response Strength of Passive Full-Order Exceptional Points}
\author{Jan Wiersig}
\affiliation{Otto von Guericke University Magdeburg, Faculty of Natural Sciences, Institute of Physics, Universit{\"a}tsplatz 2, 39106 Magdeburg, Germany}
\email{jan.wiersig@ovgu.de}

\begin{abstract}
We consider finite-dimensional non-Hermitian systems in which all modes coalesce at a single exceptional point. For a rank-one anti-Hermitian contribution, the spectral response strength is computed exactly. For passive systems, we then derive a sharp upper bound at fixed exceptional-point decay rate and show that it is reached only when the \HL{decay operator has rank one, corresponding to a single independent dissipative direction.} This establishes rank-one loss as the optimal configuration for enhancing the response of a full-order exceptional point. The bound holds for arbitrary order and improves on earlier passive estimates beyond second order. Two third-order trimers illustrate the result: a rank-one-loss system saturates the bound, whereas a passive rank-two-loss realization remains below it. The result turns dissipation into a design resource: concentrating the loss into a single \HL{effective dissipative direction maximizes the
highest-order spectral response of a passive full-order} exceptional-point device.
\end{abstract}

\maketitle

\section{Introduction}
Exceptional points (EPs) are non-Hermitian degeneracies at which eigenvalues and the corresponding eigenvectors coalesce. At an EP of order $n$, the Hamiltonian possesses a Jordan chain of length $n$, and the resolvent contains poles up to order $n$. A generic perturbation therefore produces an eigenvalue splitting proportional to the $n$th root of the perturbation strength~\cite{Kato66}, which, for sufficiently small perturbations, is significantly larger than for conventional degeneracies (diabolic points).  The prefactor of this fractional-power law is not universal. It depends on the nonnormal structure of the Hamiltonian~\cite{TE05} and can be characterized by the spectral response strength~\cite{Wiersig22}.

EPs have been investigated in a broad range of wave and quantum systems, with particularly extensive activity in optics and photonics~\cite{FE17,MA19,ORN19}. Proposed and demonstrated applications include mode selection~\cite{HMH14} and directional lasing~\cite{POL16}, asymmetric state conversion~\cite{XMJ16,DMB16}, coherent perfect absorption~\cite{SHR19}, control of light transport~\cite{XLJ23}, non-Hermitian topology~\cite{BBK21}, and sensing~\cite{Wiersig14b,COZ17,HHW17,Wiersig26}. 

Rank-one loss operators arise naturally when a finite set of internal states couples coherently to a single external decay channel. In projection-operator and effective-Hamiltonian descriptions, the anti-Hermitian part of the Hamiltonian represents decay and is given by a sum of outer products of the channel-coupling amplitudes; the associated decay matrix therefore has rank no greater than the number of open channels~\cite{MW69,SZ89,Rotter09}. Hence, a single open channel yields a rank-one anti-Hermitian part, or equivalently, a rank-one decay matrix. \HL{Conversely, several physical decay channels can also produce a rank-one decay matrix if their coupling vectors are linearly dependent, in which case they constitute a single independent dissipative direction. This rank-one dissipative structure} is relevant, for example, to nuclear and mesoscopic resonances with only one energetically open reaction channel, to wave resonators coupled to a single effective port, and to collective emission through one common reservoir channel. The same factorized structure underlies superradiant width segregation in open quantum systems~\cite{AZ11}. The corresponding loss vector is generally a coherent superposition of internal modes and need not coincide with one basis state.

The spectral response strength~$\xi$ is the norm of the highest-order pole coefficient of the Green's function and is a measure of the eigenvalue splitting under perturbation~\cite{Wiersig22}. Later work connected it to the distance from diabolic points, Petermann factors, phase rigidities, and EPs embedded in larger Hilbert spaces~\cite{Wiersig22b,Wiersig23,Wiersig23b,HS24,KW25,KWS25}. For full-order EPs, where the EP order equals the Hilbert-space dimension, Ref.~\cite{Wiersig22b} established the following upper bound on the spectral response strength for passive systems (i.e., systems with loss but no gain):
\begin{equation}\label{eq:rcapassve}
\xi \leq \left(\sqrt{2n}\,|\im\EP|\right)^{n-1} 
\end{equation}
where $n$ is the order and $\EP$  the eigenvalue of the EP.  The bound is not sharp for $n>2$.  For $n = 3$, numerical results in Ref.~\cite{Wiersig22} suggested the sharp upper bound
\begin{equation}\label{eq:rcapassve3}
	\xi \leq (2|\im\EP|)^2 .
\end{equation}
Equality holds for this bound, as well as for inequality~(\ref{eq:rcapassve}) in the case $n=2$, whenever the decay operator has rank one. Thus, rank-one losses maximize the spectral response strength for a fixed EP decay rate $2|\im\EP|$, which is potentially useful for sensing applications~\cite{WR26}.

% aim
In this paper, we prove analytically for full-order EPs in passive systems the sharp bound
\begin{equation}
\xi\leq \left(2|\im\EP|\right)^{n-1}
\end{equation}
which is strictly tighter than Eq.~\eqref{eq:rcapassve} for $n>2$ and proves the bound previously suggested by numerical results in Eq.~\eqref{eq:rcapassve3} for $n=3$. We further show that, within the class of passive full-order EPs, equality is attained if and only if the decay operator has rank one. Independently, the exact response strength for a rank-one anti-Hermitian contribution is valid without assuming passivity and applies to either loss or gain.

The outline of the paper is as follows. Section~\ref{sec:model} introduces the model and underlying assumptions, while Sec.~\ref{sec:response} reviews the concept of spectral response strength.  In Sec.~\ref{sec:rankone}, the matrix determinant lemma gives the exact rank-one response for either loss or gain without assuming passivity. Passivity is imposed in Sec.~\ref{sec:optimality}, where a Schur decomposition and the correlation-matrix structure of the normalized decay operator yield the sharp upper bound, its equality condition, and an explicit saturating family for every order. \HL{Section~\ref{sec:consequences} compares the resulting bound with the earlier passive estimate and discusses its perturbative consequences.} Section~\ref{sec:examples} compares a passive rank-one-loss trimer with a passive version of the parity-time-symmetric trimer with rank-two loss. Section~\ref{sec:conclusion} concludes the paper.

\paragraph*{Use of an AI tool.}
OpenAI's ChatGPT, with the GPT-5.6 Pro model, was used as an interactive aid in examining intermediate algebraic steps. The author independently checked all derivations and cited sources before inclusion.

\section{Model and assumptions}
\label{sec:model}
Consider the non-Hermitian Hamiltonian
\begin{align}
 \Ham&=\Herm-\frac{\ii}{2}\Loss,
 \label{eq:H}\\
 \Herm&=\Herm^\dagger,
 \qquad
 \Loss=\Loss^\dagger.
 \label{eq:general_decomposition}
\end{align}
The Hermitian operator $\Loss$ is related to the anti-Hermitian part of $\Ham$,
\begin{equation}
 \Loss=\ii(\Ham-\Ham^\dagger).
 \label{eq:Gamma_def}
\end{equation}
No positivity or passivity condition is imposed at this stage. Positive and negative eigenvalues of the decay matrix $\Loss$ describe loss and gain contributions, respectively.

We assume that the complete $n$-dimensional state space, with $n\geq2$, is the generalized eigenspace of one EP of order $n$. Equivalently,
\begin{equation}
 \det(E\id-\Ham)=(E-\EP)^n,
 \label{eq:charpoly}
\end{equation}
with geometric multiplicity one. The shifted Hamiltonian
\begin{equation}
 \Nil:=\Ham-\EP\id
 \label{eq:Ndef}
\end{equation}
is nilpotent with index $n$,
\begin{equation}
 \Nil^n=0,
 \qquad
 \Nil^{n-1}\neq0.
 \label{eq:nilpotent}
\end{equation}

It is convenient to introduce the signed linewidth parameter
\begin{equation}
 \beta:=-2\im\EP.
 \label{eq:beta}
\end{equation}
Thus $\beta>0$ corresponds to decay, whereas $\beta<0$ corresponds to amplification of the coalesced mode. Since a nilpotent matrix $\Nil$ is traceless~\cite{HJ13}, $\tr\Ham=n\EP$. Taking imaginary parts and using Eq.~\eqref{eq:H} gives the general trace relation
\begin{equation}
 \tr\Loss=n\beta.
 \label{eq:trace_relation_general}
\end{equation}

The rank-one case is obtained by imposing
\begin{equation}
 \Loss=\gamma|v\rangle\langle v|,
 \qquad
 \langle v|v\rangle=1,
 \qquad
 \gamma\in\mathbb{R}\setminus\{0\}.
 \label{eq:rank_one_loss}
\end{equation}
For $\gamma>0$ the rank-one term represents loss, while $\gamma<0$ represents gain. Equation~\eqref{eq:trace_relation_general} reduces to
\begin{equation}
 \gamma=n\beta,
 \qquad
 \im\EP=-\frac{\gamma}{2n}.
 \label{eq:trace_relation_rank_one}
\end{equation}
The analysis in Secs.~\ref{sec:model}-\ref{sec:rankone} does not require passivity. The additional condition of $\Loss$ being positive semidefinite, i.e., $\Loss\succeq0$, is introduced only in Sec.~\ref{sec:optimality}.

\section{Spectral response strength}
\label{sec:response}
For the complex-valued energy $E$, the Green's function of the Hamiltonian $\Ham$ is
\begin{equation}
 \Green(E):=(E\id-\Ham)^{-1}.
 \label{eq:Green}
\end{equation}
Using $\Ham=\EP\id+\Nil$ [Eq.~(\ref{eq:Ndef})] and the nilpotency of $\Nil$, its Laurent expansion terminates exactly~\cite{Kato66,Heiss15,Wiersig22,HKC22},
\begin{align}
 \Green(E)
 &=\frac{\id}{E-\EP}
 +\frac{\Nil}{(E-\EP)^2}
 +\cdots
 +\frac{\Nil^{n-1}}{(E-\EP)^n}.
 \label{eq:resolvent_expansion}
\end{align}
The coefficient of the highest-order pole is therefore $\Nil^{n-1}$. The spectral response strength is defined as~\cite{Wiersig22}
\begin{equation}
 \xi:=\norm{\Nil^{n-1}}_2,
 \label{eq:xi_definition}
\end{equation}
where $\norm{\cdot}_2$ denotes the spectral matrix norm~\cite{HJ13}. The range of $\Nil^{n-1}$ lies in the one-dimensional eigenspace $\ker\Nil$, and $\Nil^{n-1}\neq0$. Hence $\Nil^{n-1}$ has rank one.

For a perturbed Hamiltonian 
\begin{equation}
 \Ham(\varepsilon)=\Ham+\varepsilon\Pert,
 \qquad \varepsilon>0,
 \label{eq:perturbed_H}
\end{equation}
with generic perturbation $\Pert$, 
the leading-pole analysis gives the eigenvalue-displacement bound
\begin{equation}
 |E_j(\varepsilon)-\EP|^n
 \leq
 \varepsilon\norm{\Pert}_2\,\xi,
 \label{eq:Wiersig_bound}
\end{equation}
where higher orders in $\varepsilon$ are neglected~\cite{Wiersig22}. The quantity $\xi$ is a property of the unperturbed EP and is independent of the perturbation strength and of the chosen orthonormal basis.

\section{Exact rank-one response strength}
\label{sec:rankone}
\subsection{Determinant identity}
We now specialize to the rank-one Hermitian operator $\hat\Gamma$ in Eq.~\eqref{eq:rank_one_loss}. The matrix determinant lemma~\cite{Harville1997} states that, for an invertible matrix $\hat A$ and vectors $|u\rangle$ and $|w\rangle$,
\begin{equation}
 \det\!\left(\hat A+|u\rangle\langle w|\right)
 =\det(\hat A)
 \left(1+\langle w|\hat A^{-1}|u\rangle\right).
 \label{eq:MDL}
\end{equation}
%The identity follows by factoring out $\hat A$ and applying Sylvester's determinant theorem to the remaining rank-one update.
For Eq.~\eqref{eq:rank_one_loss},
\begin{equation}
 \Ham^\dagger=\Ham+\ii\gamma|v\rangle\langle v|,
\end{equation}
and therefore
\begin{equation}
 E\id-\Ham^\dagger
 =E\id-\Ham-\ii\gamma|v\rangle\langle v|.
\end{equation}
Apply Eq.~\eqref{eq:MDL} with
\begin{equation}
 \hat A=E\id-\Ham,
 \qquad
 |u\rangle=-\ii\gamma|v\rangle,
 \qquad
 \langle w|=\langle v|.
\end{equation}
Introducing the Green's-function matrix element in the rank-one channel,
\begin{equation}
 g(E):=\mel{v}{(E\id-\Ham)^{-1}}{v},
 \label{eq:g_def}
\end{equation}
one obtains
\begin{equation}
 1-\ii\gamma g(E)
 =\frac{\det(E\id-\Ham^\dagger)}{\det(E\id-\Ham)}.
 \label{eq:det_ratio}
\end{equation}
The identity is first established away from the spectrum, where $E\id-\Ham$ is invertible, and then interpreted as an equality of rational functions.

At the full-order EP of order $n$ ($\mathrm{EP}_n$), Eq.~\eqref{eq:charpoly} fixes the denominator. The eigenvalues of $\Ham^\dagger$ are the complex conjugates of those of $\Ham$, so that
\begin{equation}
 \det(E\id-\Ham^\dagger)=(E-\EP^*)^n.
\end{equation}
Since $\EP^*=\EP+\ii\beta$, Eq.~\eqref{eq:det_ratio} becomes
\begin{align}
 1-\ii\gamma g(E)
 &=\left(\frac{E-\EP^*}{E-\EP}\right)^n
 \nonumber\\
 &=\left(1-\frac{\ii\beta}{E-\EP}\right)^n.
 \label{eq:g_ratio}
\end{align}
Solving for $g$ gives
\begin{equation}
 g(E)=\frac{1}{\ii\gamma}
 \left[
 1-\left(1-\frac{\ii\beta}{E-\EP}\right)^n
 \right].
 \label{eq:g_exact}
\end{equation}

\subsection{Highest-order pole}
Expanding Eq.~\eqref{eq:g_exact} yields
\begin{equation}
 g(E)=-\frac{1}{\ii\gamma}
 \sum_{k=1}^{n}\binom{n}{k}
 \frac{(-\ii\beta)^k}{(E-\EP)^k}.
 \label{eq:g_binomial}
\end{equation}
The coefficient of the highest-order pole is
\begin{equation}
 -\frac{(-\ii\beta)^n}{\ii\gamma}.
 \label{eq:highest_channel_coefficient}
\end{equation}
On the other hand, the rank-one-channel matrix element of Eq.~\eqref{eq:resolvent_expansion} shows that the same coefficient is $\mel{v}{\Nil^{n-1}}{v}$. Hence
\begin{equation}
 \left|\mel{v}{\Nil^{n-1}}{v}\right|
 =\frac{|\beta|^n}{|\gamma|}
 =\frac{|\beta|^{n-1}}{n}.
 \label{eq:channel_coefficient}
\end{equation}
In the final step, we used Eq.~(\ref{eq:trace_relation_rank_one}).

Let $|R\rangle$ and $|L\rangle$ denote normalized right and left EP eigenvectors,
\begin{align}
 \Ham|R\rangle&=\EP|R\rangle,
 &\langle R|R\rangle&=1,
 \nonumber\\
 \Ham^\dagger|L\rangle&=\EP^*|L\rangle,
 &\langle L|L\rangle&=1.
 \label{eq:left_right}
\end{align}
Since $\Nil^{n-1}$ has rank one, its right and left singular directions are fixed by these eigenvectors. It can therefore be written as~\cite{KWS25}
\begin{equation}
 \Nil^{n-1}=\xi e^{\ii\phi}|R\rangle\langle L|
 \label{eq:outer_product}
\end{equation}
with a real phase $\phi$. The eigenvalue equations~(\ref{eq:left_right}) and Eqs.~\eqref{eq:Gamma_def} and~\eqref{eq:rank_one_loss} give
\begin{equation}
 \im\EP
 =\im\langle R|\Ham|R\rangle
 =-\frac{\gamma}{2}|\langle v|R\rangle|^2,
\end{equation}
and therefore with Eq.~\eqref{eq:trace_relation_rank_one}
\begin{equation}
 |\langle v|R\rangle|^2
 =\frac{\beta}{\gamma}
 =\frac{|\beta|}{|\gamma|}
 =\frac{1}{n}.
 \label{eq:right_overlap}
\end{equation}
The corresponding argument for $\Ham^\dagger|L\rangle=\EP^*|L\rangle$ gives
\begin{equation}
 |\langle v|L\rangle|^2=\frac{1}{n}.
 \label{eq:left_overlap}
\end{equation}
Equation~\eqref{eq:outer_product} then gives
\begin{equation}
 \left|\mel{v}{\Nil^{n-1}}{v}\right|
 =\xi|\langle v|R\rangle|\,|\langle L|v\rangle|
 =\frac{\xi}{n}.
 \label{eq:xi_channel_relation}
\end{equation}
Comparison with Eq.~\eqref{eq:channel_coefficient} gives
\begin{equation}
 \xi=|\beta|^{n-1}
 =\left(\frac{|\gamma|}{n}\right)^{n-1}
 =\left(2|\im\EP|\right)^{n-1}.
 \label{eq:rank_one_result}
\end{equation}
Equation~\eqref{eq:rank_one_result} is the first central result of this work. It extends the analytical result for $n=2$ and the numerical finding for $n=3$ from Ref.~\cite{Wiersig22} to arbitrary order $n$. For systems with a rank-one anti-Hermitian contribution, it provides a remarkably simple and useful way to compute the spectral response strength through a trace operation; see Eqs.~\eqref{eq:beta}-\eqref{eq:trace_relation_general}.

\section{Optimality among passive systems}
\label{sec:optimality}
The rank-one result in Eq.~\eqref{eq:rank_one_result} can be placed into a more general variational context. From this section onward, we impose passivity~\cite{Wiersig16,Wiersig19,HS22},
\begin{equation}
 \Loss\succeq0.
 \label{eq:passivity}
\end{equation}
Equation~\eqref{eq:trace_relation_general} then implies $\beta\geq0$. At a nontrivial full-order EP one has $\beta>0$, because $\beta=0$ would imply $\Loss=0$ and hence a Hermitian, nondefective Hamiltonian. In the following, $\Loss$ is an arbitrary positive-semidefinite decay operator and the EP decay rate $\beta$ is kept fixed.

By the Schur theorem~\cite{HJ13}, there exists a unitary matrix $\Unitary$ such that
\begin{equation}
 \Unitary^\dagger\Ham\Unitary=\EP\id+\SchurT,
 \label{eq:Schur}
\end{equation}
where $\SchurT$ is strictly upper triangular. Unitary invariance of the spectral norm gives
\begin{equation}
 \xi=\norm{\SchurT^{n-1}}_2.
 \label{eq:xi_Schur}
\end{equation}
Define the first-superdiagonal elements
\begin{equation}
 t_j:=(\SchurT)_{j,j+1},
 \qquad j=1,\ldots,n-1.
 \label{eq:tj}
\end{equation}
For a strictly upper-triangular $n\times n$ matrix $\SchurT$, the power $\SchurT^{n-1}$ has only one potentially nonzero matrix element,
\begin{equation}
 \SchurT^{n-1}
 =\left(\prod_{j=1}^{n-1}t_j\right)
 |e_1\rangle\langle e_n|,
 \label{eq:T_power}
\end{equation}
and therefore with Eq.~\eqref{eq:xi_Schur}
\begin{equation}
 \xi=\prod_{j=1}^{n-1}|t_j|.
 \label{eq:xi_product}
\end{equation}
The full-order EP condition $\SchurT^{n-1}\neq0$ implies that none of the factors vanishes.

Transforming the decay operator~(\ref{eq:Gamma_def}) to the same basis gives
\begin{align}
 \Loss_{\mathrm S}
 &:=\Unitary^\dagger\Loss\Unitary
 \nonumber\\
 &=\beta\id+\ii(\SchurT-\SchurT^\dagger).
 \label{eq:Gamma_S}
\end{align}
In particular, every diagonal element of $\Loss_{\mathrm S}$ equals $\beta\neq 0$. The normalized matrix
\begin{equation}
 \Corr:=\frac{\Loss_{\mathrm S}}{\beta}
 \label{eq:C_def}
\end{equation}
is positive semidefinite and has unit diagonal. It is therefore a correlation matrix~\cite{LiTam1994}. Equivalently, it can be written as a Gram matrix of unit vectors $|u_j\rangle$~\cite{HJ13},
\begin{equation}
 (\Corr)_{jk}=\langle u_j|u_k\rangle,
 \qquad
 \langle u_j|u_j\rangle=1.
 \label{eq:Gram}
\end{equation}
For neighboring indices, Eq.~\eqref{eq:Gamma_S} gives
\begin{equation}
 (\Corr)_{j,j+1}=\frac{\ii}{\beta}t_j,
 \qquad
 |t_j|=\beta|\langle u_j|u_{j+1}\rangle|.
 \label{eq:t_correlation}
\end{equation}
Substitution into Eq.~\eqref{eq:xi_product} yields the important identity
\begin{equation}
 \frac{\xi}{\beta^{n-1}}
 =\prod_{j=1}^{n-1}
 |\langle u_j|u_{j+1}\rangle|.
 \label{eq:correlation_product}
\end{equation}
Each factor on the right-hand side is at most one by the Cauchy--Schwarz inequality. Hence every passive full-order $\mathrm{EP}_n$ satisfies
\begin{equation}
 \xi\leq\beta^{n-1}.
 \label{eq:sharp_bound}
\end{equation}

The equality condition is equally restrictive. Equality in Eq.~\eqref{eq:sharp_bound} requires
\begin{equation}
 |\langle u_j|u_{j+1}\rangle|=1
 \qquad
 \text{for all }j.
 \label{eq:equality_factors}
\end{equation}
Equality in the Cauchy--Schwarz inequality implies that every neighboring pair is proportional. Therefore all Gram vectors are collinear,
\begin{equation}
 |u_1\rangle\parallel|u_2\rangle\parallel\cdots\parallel|u_n\rangle,
\end{equation}
and $\Corr$ has rank one~\cite{HJ13}. Since unitary transformations and multiplication by a nonzero scalar preserve rank,
\begin{equation}
 \rank\Corr=1
 \quad\Longleftrightarrow\quad
 \rank\Loss=1.
\end{equation}
Conversely, a rank-one correlation matrix is the Gram matrix of collinear unit vectors, so every factor on the right-hand side of Eq.~\eqref{eq:correlation_product} equals one. We have thus established the sharp statement
\begin{equation}
 \xi\leq\beta^{n-1},
 \qquad
 \xi=\beta^{n-1}
 \quad\Longleftrightarrow\quad
 \rank\Loss=1,
 \label{eq:optimality_theorem}
\end{equation}
which is the second central result of this work.

We now show that the bound is attainable for arbitrary order by an explicit construction. Choose arbitrary real phases $\theta_j$ and define, in the Schur basis,
\begin{equation}
 |q\rangle=\sum_{j=1}^{n}e^{\ii\theta_j}|e_j\rangle,
 \qquad
 |v_{\mathrm S}\rangle=\frac{|q\rangle}{\sqrt{n}},
 \label{eq:construction_vectors}
\end{equation}
together with the rank-one decay matrix
\begin{equation}
 \Loss_{\mathrm S}=\beta|q\rangle\langle q|
 =n\beta|v_{\mathrm S}\rangle\langle v_{\mathrm S}|.
 \label{eq:construction_loss}
\end{equation}
Let the strictly upper-triangular operator $\SchurT$ have matrix elements
\begin{equation}
 (\SchurT)_{jk}=
 \begin{cases}
 -\ii\beta e^{\ii(\theta_j-\theta_k)},&j<k,\\
 0,&j\geq k.
 \end{cases}
 \label{eq:construction_T}
\end{equation}
Then
\begin{equation}
 \Loss_{\mathrm S}=\beta\id+\ii(\SchurT-\SchurT^\dagger),
 \label{eq:construction_identity}
\end{equation}
and every first-superdiagonal element of $\SchurT$ has modulus~$\beta$. Consequently, $\SchurT^n=0$, $\SchurT^{n-1}\neq0$, and
\begin{equation}
 \norm{\SchurT^{n-1}}_2=\beta^{n-1}.
 \label{eq:construction_xi}
\end{equation}
For any $E_0\in\mathbb{R}$, the Hamiltonian
\begin{equation}
 \Ham_{\mathrm S}=\left(E_0-\frac{\ii\beta}{2}\right)\id+\SchurT
 \label{eq:construction_H}
\end{equation}
therefore realizes a passive full-order $\mathrm{EP}_n$ with rank-one loss and saturates Eq.~\eqref{eq:sharp_bound}.

At fixed total loss $\gamma_{\mathrm{tot}}:=\tr\Loss=n\beta$, the maximal response is
\begin{equation}
 \xi_{\max}
 =\left(\frac{\gamma_{\mathrm{tot}}}{n}\right)^{n-1}.
 \label{eq:xi_max_total_loss}
\end{equation}
Equation~\eqref{eq:rank_one_result} and the construction above both saturate this bound. Within the class of passive full-order EP$_n$ Hamiltonians at fixed $\beta$, the maximizing decay operators are precisely those of rank one.

\HL{
\section{Comparison with earlier passive bounds and perturbative consequences}
\label{sec:consequences}
The sharp bound derived in Sec.~\ref{sec:optimality} can now be compared with the earlier arbitrary-order estimate and translated into bounds on the perturbative eigenvalue displacement.}
The distance-based analysis of Ref.~\cite{Wiersig22b} gave the arbitrary-order estimate for passive systems
\begin{equation}
 \xi\leq
 \left(\sqrt{2n}\,|\im\EP|\right)^{n-1}
 =\left(\sqrt{\frac{n}{2}}\,\beta\right)^{n-1}.
 \label{eq:previous_passive_bound}
\end{equation}
The ratio of the right-hand side of Eq.~\eqref{eq:previous_passive_bound} to the sharp right-hand side in Eq.~\eqref{eq:sharp_bound} is
\begin{equation}
 \left(\frac{n}{2}\right)^{(n-1)/2}.
 \label{eq:bound_improvement_factor}
\end{equation}
The bounds agree for $n=2$, whereas Eq.~\eqref{eq:sharp_bound} is strictly stronger for every $n>2$. The improvement follows from retaining the positive-semidefinite correlation structure of the complete decay operator rather than bounding it only through a global matrix norm. 
Note that for $n=3$, Eq.~\eqref{eq:sharp_bound} is fully consistent with the numerically observed bound in Ref.~\cite{Wiersig22}. 

Combining Eqs.~\eqref{eq:Wiersig_bound} and \eqref{eq:sharp_bound} gives the universal passive-system estimate
\begin{equation}
 |E_j(\varepsilon)-\EP|^n
 \leq
 \varepsilon\norm{\Pert}_2\,\beta^{n-1}.
 \label{eq:universal_splitting_bound}
\end{equation}
For a fixed total loss, this can be written as
\begin{equation}
 |E_j(\varepsilon)-\EP|^n
 \leq
 \varepsilon\norm{\Pert}_2
 \left(\frac{\gamma_{\mathrm{tot}}}{n}\right)^{n-1}.
 \label{eq:universal_splitting_total}
\end{equation}
The characteristic displacement scale obeys
\begin{equation}
 |E_j(\varepsilon)-\EP|
 \lesssim
 \left(\varepsilon\norm{\Pert}_2\right)^{1/n}
 \beta^{(n-1)/n}.
 \label{eq:splitting_scale}
\end{equation}
The largest possible prefactor is realized by rank-one loss. A particular perturbation need not saturate Eq.~\eqref{eq:universal_splitting_bound}, because the bound also depends on the orientation of $\Pert$ relative to the highest-order pole coefficient~\cite{Wiersig22,Wiersig22c}.

%For $n=2$ and $n=3$, Eq.~\eqref{eq:optimality_theorem} gives
%\begin{equation}
% \xi_{\max}=\beta,
 %\qquad
 %\xi_{\max}=\beta^2,
%\end{equation}
%respectively.

\section{Two examples}
\label{sec:examples}
We compare two passive trimers at a full-order $\mathrm{EP}_3$. The examples are chosen to have the same EP decay parameter $\beta$, so that the optimality result can be tested directly.

\subsection{Rank-one-loss trimer}
We first consider a trimer with rank-one loss, previously studied in the context of atomic Bose-Einstein condensates~\cite{WLX24} and quantum Fisher information~\cite{WR26}. Specifically, we consider a real nearest-neighbor coupling matrix with loss acting on the first basis state,
\begin{align}
 \Herm&=
 \begin{pmatrix}
 0&V_1&0\\
 V_1&0&V_2\\
 0&V_2&0
 \end{pmatrix},
 &
 \Loss&=\gamma|e_1\rangle\langle e_1|.
 \label{eq:rankone_example_general}
\end{align}
The two couplings can be tuned such that all three eigenvalues of $\Ham=\Herm-\ii\Loss/2$ coalesce at $\EP=-\ii\gamma/6$. One convenient choice is
\begin{equation}
 V_1=\frac{\sqrt{6}\,\gamma}{9},
 \qquad
 V_2=\frac{\gamma}{6\sqrt{3}}.
 \label{eq:rankone_example_couplings}
\end{equation}
The shifted operator $\Nil=\Ham-\EP\id$ then satisfies $\Nil^3=0$ and $\Nil^2\neq0$. Since $\beta=\gamma/3$, a straightforward calculation gives the rank-one result 
\begin{equation}
 \xi_{\mathrm{r1}}=\norm{\Nil^2}_2=\beta^2.
 \label{eq:rankone_example_xi}
\end{equation}
Thus this trimer saturates the passive upper bound~(\ref{eq:optimality_theorem}). Its detailed eigenvectors and Jordan chain are not needed: saturation follows solely from the full-order-EP condition and $\rank\Loss=1$.

\subsection{Passive form of the parity-time-symmetric trimer}
A standard parity-time-symmetric trimer consists of three equally coupled modes with balanced gain and loss on the outer sites~\cite{DG12},
\begin{equation}
 \Ham_{\mathcal{PT}}=
 \begin{pmatrix}
 \ii g&J&0\\
 J&0&J\\
 0&J&-\ii g
 \end{pmatrix}.
 \label{eq:PT_trimer}
\end{equation}
At $g=\sqrt{2}J$, this matrix has a third-order EP with eigenvalue zero. This model underlies the higher-order EP sensing experiment of Hodaei \textit{et al.}~\cite{HHW17}. Adding a uniform loss $-\ii g\id$ leaves the nilpotent part, the EP order, and the spectral response strength unchanged, but moves the coalesced eigenvalue to $\EP=-\ii g$:
\begin{equation}
 \Ham_{\mathrm p}=\Ham_{\mathcal{PT}}-\ii g\id
 =\Herm_{\mathrm p}-\frac{\ii}{2}\Loss_{\mathrm p}.
 \label{eq:passive_PT_shift}
\end{equation}
The decay operator is
\begin{equation}
 \Loss_{\mathrm p}=
 \begin{pmatrix}
 0&0&0\\
 0&2g&0\\
 0&0&4g
 \end{pmatrix},
 \qquad
 \rank\Loss_{\mathrm p}=2,
 \label{eq:passive_PT_loss}
\end{equation}
so the shifted model is passive but not a rank-one-loss system. Its EP decay parameter is, according to Eq.~\eqref{eq:beta},  $\beta=2g$. At $g=\sqrt{2}J$, direct evaluation of the highest nilpotent power gives (see also Ref.~\cite{Wiersig22})
\begin{equation}
 \xi_{\mathrm{r2}}
 =\norm{(\Ham_{\mathrm p}-\EP\id)^2}_2
 = \norm{\Ham_{\mathcal{PT}}^2}_2
 =4J^2
 =\frac{\beta^2}{2}.
 \label{eq:ranktwo_example_xi}
\end{equation}
The numerical factor $1/2$ is specific to this trimer and is not determined by the loss rank. For fixed $\beta$ and loss rank larger than one, the response strength depends on the full correlation structure in Eq.~\eqref{eq:correlation_product}; rank one uniquely characterizes saturation of the upper bound, but the rank alone does not determine the response below the bound.

The comparison of the rank-one and rank-two models is therefore immediate:
\begin{equation}
 \xi_{\mathrm{r1}}=\beta^2,
 \qquad
 \xi_{\mathrm{r2}}=\frac{\beta^2}{2}.
 \label{eq:example_comparison}
\end{equation}
Both systems are passive full-order $\mathrm{EP}_3$ models at the same decay rate $\beta$. The rank-one example reaches the universal maximum in Eq.~\eqref{eq:sharp_bound}, whereas the passive parity-time trimer, whose \HL{decay operator has two independent dissipative directions}, falls below it by a factor of two. The examples illustrate that the EP order and decay rate alone do not determine the response strength; the rank and correlation structure of the loss operator are decisive.

\section{Conclusion}
\label{sec:conclusion}
We have analyzed $n\times n$ Hamiltonians hosting an EP of order $n$. For a rank-one anti-Hermitian part, the matrix determinant lemma yields a simple exact expression for the spectral response strength, applicable to both rank-one loss and rank-one gain.

After imposing passivity, a unitary Schur decomposition shows that the normalized decay operator is a correlation matrix and that the response strength is a product of adjacent correlation magnitudes. The Cauchy-Schwarz inequality then yields the sharp arbitrary-order bound, with equality precisely for rank-one loss. An explicit construction establishes attainability for every order. Compared with the earlier distance-based passive estimate, the new bound removes a factor $(n/2)^{(n-1)/2}$ from the upper bound for $n>2$. The two $\mathrm{EP}_3$ examples emphasize the equality condition: rank-one loss reaches the bound, while the passive version of the parity-time-symmetric trimer with rank-two loss remains below it.

The optimality proof assumes that the EP order equals the dimension of the Hilbert space. Physically, all available modes then belong to a single coalescing Jordan chain, and the highest-order response is built up coherently along one unique sequence of mode couplings. If an $\mathrm{EP}_n$ is embedded in a larger space (see, e.g., Refs.~\cite{Wiersig23b,KWS25}), noncoalescing spectator modes can provide additional pathways for storing, redistributing, or dissipating the excitation. The highest-order contribution of the Green's function then contains additional spectral projectors, and the simple product formula in Eq.~\eqref{eq:xi_product} no longer applies directly.

From a physical perspective, the bound provides a direct design principle for passive wave and photonic systems: at a fixed EP decay rate, concentrating dissipation into a single \HL{independent coupling direction, or effective loss channel,} maximizes the highest-order spectral response, whereas \HL{introducing additional linearly independent dissipative directions} necessarily reduces it. This makes loss-channel engineering a practical resource for optimizing coupled resonators, integrated photonic circuits, and other open-wave devices operating near EPs. 
This result concerns the highest-order spectral response strength. The performance of a specific sensing scheme also depends on the perturbation, excitation, readout, and noise characteristics, see, e.g., Refs.~\cite{LC18,ZSH19,LS24,WR26}.

An interesting open direction is the extension to fragmented EPs~\cite{BS25,BS25b}, where several Jordan chains are present. The single-chain argument used here no longer applies directly, and it remains open whether analogous bounds exist and whether rank-one loss remains optimal. Establishing such bounds could connect dissipation engineering to the unconventional bulk and edge responses predicted for fragmented exceptional points.

\begin{acknowledgments}
The author used OpenAI's ChatGPT, with the GPT-5.6 Pro model, for manuscript organization, language editing, and LaTeX preparation. The author formulated and directed the prompts, critically evaluated and revised all generated material, and takes full responsibility for the content of the manuscript.
I thank J. Kullig for fruitful discussions.
%
%I acknowledge the support by the Otto von Guericke University Magdeburg.
\end{acknowledgments}

\section*{Data availability}
No research data or software were generated. All information supporting the results is contained in the article.
 
%\bibliography{../../bib/fg4,../../bib/extern}

\end{document}